# Electrostatic Control Enables Robust Helical Edge Channel Transport in III-V Quantum Spin Hall Insulators

Manuel Meyer[1,a)], Tobias Fähndrich[1,2], Sebastian Schmid[1], Justus Walter[1], Martin Kamp[3], Adriana Wolf[1], Sergey Krishtopenko[1,4], Guillaume Sigu[4], Jean-Baptiste Rodriguez[5], Eric Tournie[5,6], Gerald Bastard[1,7], Frederic Teppe[4], Fabian Hartmann[1], Sven Höfling[1] and Benoit Jouault[4,b)]

[1]*Julius-Maximilians-Universität Würzburg, Physikalisches Institut and Würzburg-Dresden Cluster of Excellence ctd.qmat, Lehrstuhl für Technische Physik, Am Hubland, 97074 Würzburg, Deutschland*

[2]*Department of Physics and Astronomy, University of British Columbia, Vancouver, British Columbia, V6T1Z4, Canada*

[3]*Physikalisches Institut and Röntgen Center for Complex Material Systems, 97074 Würzburg, Germany*

[4]*Laboratoire Charles Coulomb (L2C), UMR 5221 CNRS-Université de Montpellier, Montpellier, France.*

[5]*IES, Université de Montpellier, CNRS, F-34000 Montpellier, France*

[6]*Institut Universitaire de France, F-75005 Paris, France*

[7]*Physics Department, École Normale Supérieure, PSL 24 rue Lhomond, 75005 Paris, France*

Quantum spin Hall transport in InAs/GaInSb-based two-dimensional topological insulators can be limited by parasitic bulk and edge contributions. We demonstrate that these limitations are effectively mitigated through electrostatic control in dual-gated InAs/GaInSb/InAs trilayer quantum wells grown on AlSb quasi-substrates. In macroscopic Hall bars exceeding the phase coherence length, a multi-probe analysis reveals an insulating bulk and a constant edge resistance over a wide electric-field range. In microscopic devices with edge lengths below the phase coherence lengths, the edge resistance remains robust and quantized across a broad field range, revealing the intrinsic resilience of helical edge channels to electric-field perturbations. Only beyond a threshold value, parasitic edge contributions emerge. These results establish dual gating as a reliable strategy to suppress parasitic conduction while stabilizing helical edge transport, providing a versatile and reproducible platform for tunable topological transport in III-V quantum spin Hall systems.

a) Email: manuel.meyer@uni-wuerzburg.de
b) Email: benoit.jouault@umontpellier.fr

## I. Introduction

Among the many predicted two-dimensional topological insulators (2D TIs) [1–7], the InAs/(Ga,In)Sb material system is a promising platform for topological electronics applications. Its unique electron–hole bilayer structure, with spatially separated carriers in InAs and (Ga,In)Sb layers [8–10], enables tuning between trivial and topological phases through layer thickness engineering [11,12], electrical gating [13–15], or illumination [16–18]. Combined with the weak temperature dependence of its band ordering [19,20] it makes the system attractive for quantum spin Hall (QSH) devices that potentially operate above cryogenic temperatures [21]. In addition, the growth and processing of III–V semiconductors are highly advanced leading to several device applications [22–24], and the material system is compatible with Si-based chip technology [25], further supporting its potential for integration in topological electronics. In the ideal QSH regime, transport is confined to helical edge channels surrounding an insulating bulk [26–28], giving rise to quantized local and nonlocal resistance values determined by the measurement geometry [3,29]. However, early studies on InAs/GaSb bilayer quantum wells (BQWs) revealed two key obstacles to robust QSH transport: residual bulk and parasitic trivial edge conduction [30–34]. Both originate from intrinsic material properties, such as native p-doping of GaSb [35–37] and Fermi level pinning in InAs [38–42], and limited the observation of quantized edge transport in this system [43]. To address these limitations, strained InAs/GaInSb BQWs have been developed to enhance the hybridization gap to ~30 meV [14,44], and further improvements can be achieved using InAs/GaInSb/InAs trilayer quantum wells (TQWs), which can reach inverted gap energies above 40 meV when grown on AlSb quasi-substrates [11,45]. In such a TQW grown on AlSb, the QSHE up to elevated temperatures of T = 60 K was recently reported [21], showing the perspective of integrating TIs based on III-V semiconductors into topological electronics applications. Although various strategies have been proposed to suppress residual bulk conductivity [43,46–48], the role of parasitic edge contributions remains less well understood. These trivial edge channels, typically n-type in character, have been predicted to arise from Fermi level pinning at the InAs surface[38–42] and may be tunable or suppressible via gate biases [31].

In this work, we investigate dual-gated InAs/GaInSb/InAs TQWs grown on AlSb quasi-substrate to simultaneously mitigate parasitic bulk and edge conductions and realize electric-field independent helical edge channel transport. Monitoring edge and bulk resistivities on macroscopic Hall bar devices reveal an insulating bulk and a constant edge resistivity that, in microscopic devices, is quantized to $h/2e^2$. Above a threshold voltage, the quantized edge resistance decreases, signaling the onset of parasitic edge contributions. The constant, quantized resistance over a broad gate-voltage range demonstrates the absence of parasitic channels and the robustness of the helical edge channels against perturbations through electric fields.

## II. Dual-Gate Control of the Topological Band Structure

Figure 1 shows a cross-sectional high-resolution scanning transmission electron microscopy (STEM) image of the investigated InAs/GaInSb/InAs TQW. The active region and surrounding barriers are highlighted and color-coded according to the schematic in the bottom-right inset, which illustrates the full layer stack. A larger section of the STEM image and more information about growth and device fabrication can be found in Fig. S1 of the Supplemental Material (SM) [49–52]. The image confirms high crystalline quality, smooth and well-defined interfaces, and a symmetric TQW structure without pronounced interfacial roughness. Interface fluctuations within the active region remain below one monolayer, indicating excellent structural homogeneity relevant for device performance. The extracted layer thicknesses are consistent with the designed values, confirming precise control of the trilayer architecture. The top-right inset displays the calculated band structure at zero electric field, revealing an inverted band ordering with a finite topological gap between the E2 and E1 subbands and a band gap energy of 27 meV. More information about band structure calculations, including used band parameters [53], can be found in the SM [49].

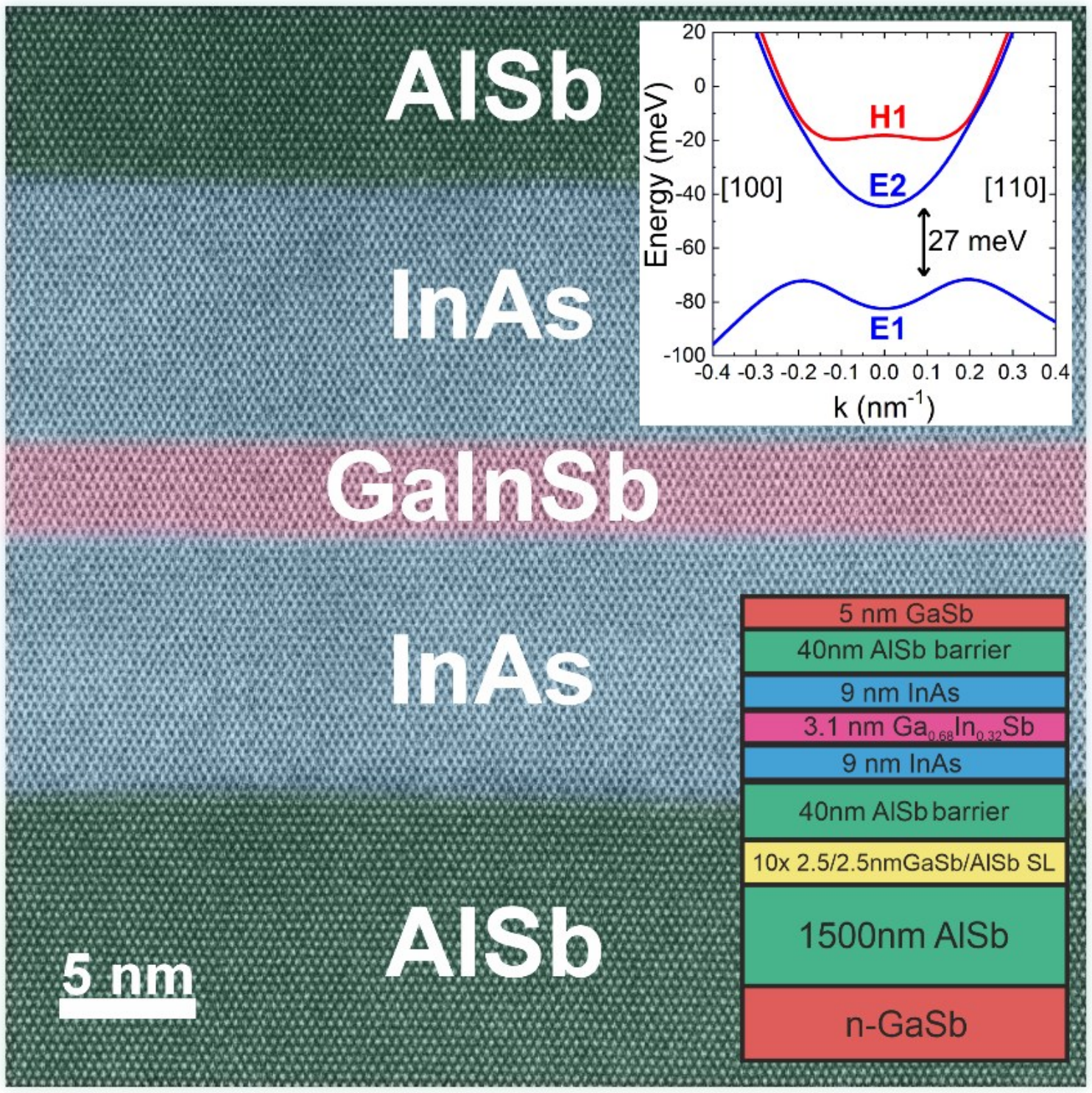

**Fig. 1.** STEM image of the investigated InAs/GaInSb/InAs TQW. The AlSb barriers and TQW layers are depicted and color-coded to match the layer schematic shown in the bottom right inset. The image reveals sharp interfaces and high structural uniformity. The top right inset displays the calculated topological band structure of the TQW, showing an inverted energy gap of $E_{gap}$ = 27 meV.

The dual gating of InAs/(Ga,In)Sb BQWs and the associated NI to TI phase transition under moderate electric fields have been studied in the literature [13,14]. However, stable back-gate operation has remained challenging due to leakage currents, which can be mitigated by growing on an electrically insulating AlSb quasi-substrate, also allowing larger band-gap energies. To explore dual-gate behavior of the TQW structure, we fabricated macroscopic Hall bar devices and investigated their transport response to top-gate ($V_{TG}$) and back-gate ($V_{BG}$) voltages, as shown in Fig. 2. Measurements were performed at T = 4.2 K with a dc current of 100 nA unless stated otherwise. Figure 2(a) displays the four-terminal resistance $R_{03,12}$ as a function of $V_{TG}$ and $V_{BG}$ for a macroscopic Hall bar device (see inset). Here, $R_{ij,kl}$ = $V_{kl}/I_{ij}$, where $V_{kl}$ is the voltage drop between contacts k and l and $I_{ij}$ is the current injected from I to j. As $V_{BG}$ increases, the resistance peak shifts towards more negative $V_{TG}$ values and decreases in magnitude from 320 kΩ to 230 kΩ. This shift reflects the need for more negative $V_{TG}$ to set the Fermi level within the band gap at higher $V_{BG}$. From the slope of the peak shifts, three regions (I-III) with different effective back-gate capacitances are identified, consistent with a simple capacitor model for the TQW[13]. While the top-gate capacitance remains constant ($C_{TG} \approx 48$ nF/cm$^2$), as it is separated by an insulating oxide layer from the TQW, the effective back-gate capacitance can vary due to the semiconductor separation, making it field-dependent [54]. Extracted back-gate capacitances are: $C_{1,BG} \approx 15$ nF/cm$^2$, $C_{2,BG} \approx 6$ nF/cm$^2$, and $C_{3,BG} \approx 22$ nF/cm$^2$, which correspond to a 2DEG tunability of $dn_1/dV_{BG} \approx 0.9 \times 10^{11}$ cm$^{-2}$V$^{-1}$, $dn_2/dV_{BG} \approx 0.37 \times 10^{11}$ cm$^{-2}$V$^{-1}$ and $dn_3/dV_{BG} \approx 1.4 \times 10^{11}$ cm$^{-2}$V$^{-1}$. A detailed discussion of the capacitor model is provided in Fig. S2 in the SM [49]. These results demonstrate that substantial band-structure tuning via the back gate is achievable over at least a 20 V range without leakage.

To investigate the evolution of the band structure with applied electric fields, temperature-dependent measurements were conducted from $V_{BG}$ = +10 V to -10 V in 1 V steps. Fig. 2(b) shows the extracted maximum resistance $R_{max}$ versus inverse temperature (1/T) for three representative $V_{BG}$ values: –10 V (red), 0 V (orange), and +6 V (green), reported in Fig. 2(a). These gate voltages were selected to represent characteristic stages of the electric-field evolution of the band structure rather than one point in each of the regions of different effective capacitances. In particular, $V_{BG}$ = +6 V was chosen because it corresponds approximately to the maximum extracted topological energy gap, allowing the evolution of the gap to be illustrated most clearly. At low temperatures up to T = 10 K, $R_{max}$ is weakly temperature dependent, indicating hopping transport [55]. To extract the band gap energy $E_{gap}$, a linear fit function was applied to the high-temperature regime, where thermally activated transport dominates. $E_{gap}$ clearly varies with $V_{BG}$, reflecting changes in the thermally activated transport $R_{max} \sim \exp(E_{gap}/2k_BT)$, where $k_B$ represents the Boltzmann constant [18]. The complete evolution of $E_{gap}$ as a function of $V_{BG}$ is presented in Fig. 2(c). For $V_{BG}$ = -10 V to -6 V, the gap energy remains nearly constant at $E_{gap} \approx 28.5$ meV, as the top- and back-gate fields approximately cancel: $E_{TG} \approx V_{TG,peak}/d_{TG} \approx V_{BG,peak}/d_{BG} \approx E_{BG}$, with $V_{TG,peak}$ and $V_{BG,peak}$ corresponding to the voltage positions of the resistance peak, whereas $d_{TG}$ and $d_{BG}$ denote the distances from the top and back gate to the TQW, respectively. This near equality implies that the total perpendicular electric field across the TQW, $E_{TG}$ - $E_{BG}$, is close to zero, resulting in minimal band bending and primarily tuning of the Fermi energy within the gap.

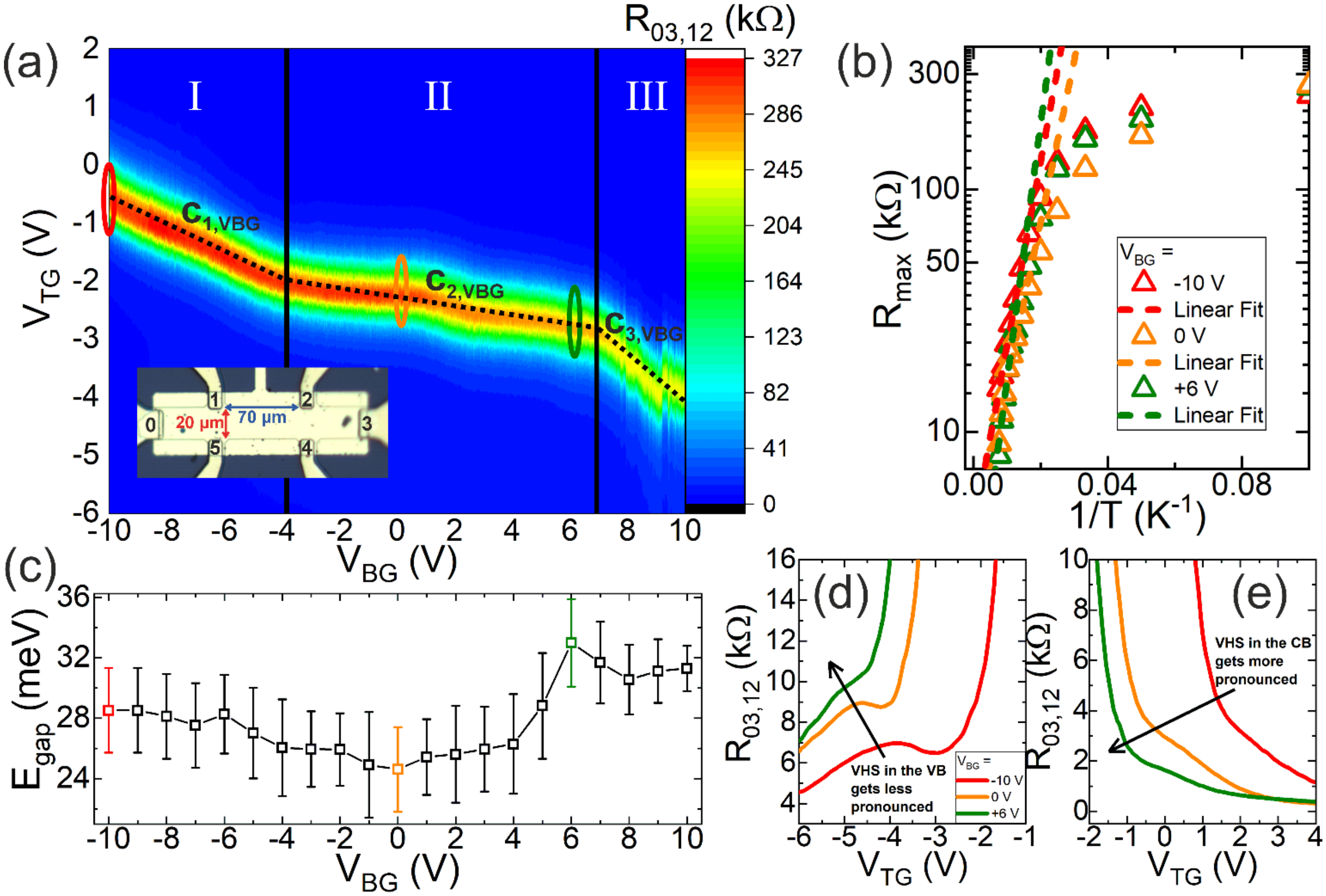

**Fig. 2.** (a) Four terminal longitudinal resistance $R_{03,12}$ as a function of $V_{BG}$ = +10 to -10 V and $V_{TG}$ = +2 to -6 V at T = 4.2 K (inset: optical image of the Hall bar). A diagonal line of maximum resistance indicates the charge neutrality condition, along which the Fermi level is positioned within the topological gap. Three regions with distinct effective back-gate capacitances are identified from the slope of the resistance peak. Colored markers indicate three representative positions for $V_{BG}$ = -10 V (red), 0 V (orange) and +6 V (green). (b) Arrhenius plots of the corresponding maximum resistance $R_{max}$ versus inverse temperature 1/T. (c) Extracted gap energies $E_{gap}$ as a function of $V_{BG}$, demonstrating electric-field tunability of the topological gap. (d),(e) Evolution of the van Hove singularities in the valence and conduction bands with increasing $V_{BG}$, indicating the electric-field tuning of the topological band structure.

As $V_{BG}$ increases from negative values towards $V_{BG} = 0$ V, $E_{gap}$ shows a slight tendency to decrease before increasing again and reaching a maximum value of $E_{gap} \approx 33.1$ meV for $V_{BG} = +6$ V. Therefore, the band gap energy can be optimized using a dual-gating approach. This tunability is further reflected in the evolution of the van Hove singularities (VHS) in the valence and conduction bands (Figs. 2(d,e)), confirming that the topological band structure responds to the applied electric field. Here, data are shown for the up-sweep, where the visibility of the VHS is better due to lower quantum scattering times [56]. A more detailed analysis of the subband evolution is provided in Fig. S3 in the SM [49].

This demonstrates how powerful the dual-gating approach is to finely tune key electronic properties. By adjusting the perpendicular electric field across the quantum well, one can modulate the band gap energy or other properties such as, for example, the crossing wave vector $k_{cross}$ between the E1 and H1 subbands. As bulk conductivity strongly depends on $E_{gap}$ [57,58], the bulk resistivity $\rho_{bulk}$ correspondingly varies with the applied electric field. The electric field control might further impact parasitic edge contributions as suggested in Ref. [31]. To quantify the evolution of bulk and edge resistivities under electric-field control, we extracted these quantities  from local and nonlocal measurements [59].

## III. Electrostatic Optimization of Helical Edge Transport

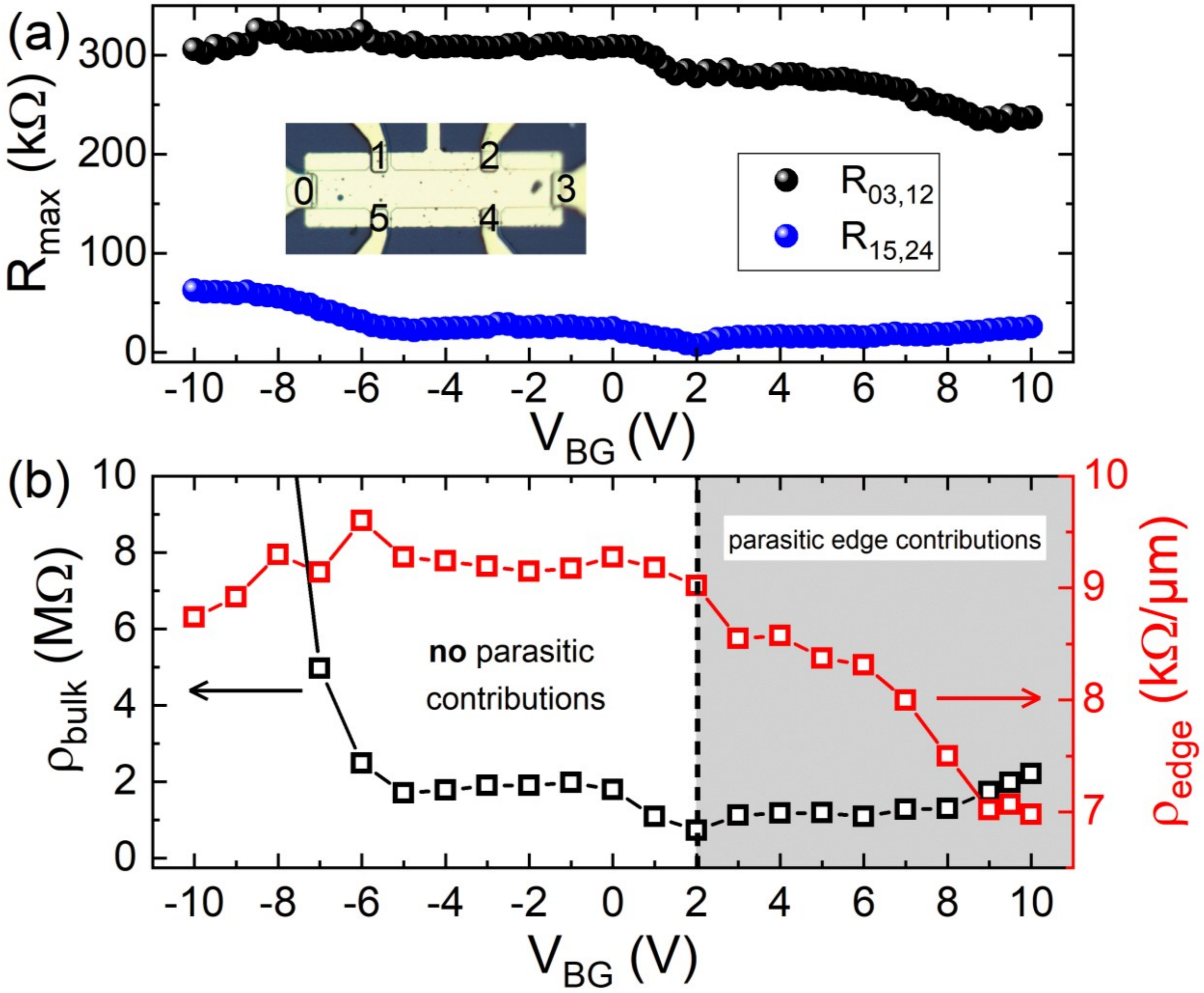

**Fig. 3.** (a) Maximum resistance $R_{max}$ as a function of $V_{BG}$ = +10 V to -10 V for $R_{03,12}$ (local configuration, black) and $R_{15,24}$ (non-local configuration, blue). (b) Extracted bulk resistivity $\rho_{bulk}$ and edge resistivity $\rho_{edge}$. $\rho_{bulk}$ remains in the MΩ-regime, evidencing an insulating bulk over the full range. Between $V_{BG}$ = -10 to +2 V, $\rho_{edge}$ is mostly constant at 9 kΩ/µm and decreases for $V_{BG}$ > +2 V, signaling the onset of parasitic edge contributions.

The corresponding maxima of local resistance $R_{03,12}$ (in black) and non-local resistance $R_{15,24}$ (in blue) are shown in Fig. 3(a). Notably, the positions of local and non-local maxima occur at identical gate voltages, confirming a consistent transport regime. Both maximum resistance values get altered with changes in the electric field. By fitting both configurations simultaneously (see Ref. [59] for more information on the analysis), we extract the bulk and edge resistivities as a function of $V_{BG}$, shown in Fig. 3(b). Over the full investigated range, the bulk resistivity remains in the MΩ regime and increases further toward negative $V_{BG}$, demonstrating a highly insulating bulk that can be optimized via electric-field tuning. This confirms that residual bulk conductivity, which is one of the central limitations of InAs/(Ga,In)Sb-based topological systems, can be effectively suppressed in strained InAs/GaInSb/InAs trilayer quantum wells. In contrast, the edge resistivity remains constant at approximately 9 kΩ/μm over a broad gate range from $V_{BG} = -10$ V to +2 V, compatible with helical edge transport with a backscattering length $\lambda \simeq 3$ μm as previously observed [9, 37]. For $V_{BG} > +2$ V, however, the edge resistivity decreases continuously. Since the bulk remains highly insulating in this regime and the inverted band structure persists across the full gate range (Fig. 2), this reduction cannot originate from bulk conduction or a collapse of the topological phase. Instead, following Nguyen et al. [31], this behavior is consistent with the gradual population of trivial n-type edge channels, likely arising from Fermi-level pinning at the lateral InAs surface [30–33], and modulated by gate biases. As these trivial edge states become increasingly occupied at more positive gate voltages, they provide an additional parallel conduction path, thereby lowering the total measured edge resistance.

The abrupt onset of this decrease further argues against an alternative explanation based on changes in the phase coherence length, which in similar systems varies continuously across the top-gate-voltage range [21,60], rather than remaining constant and then dropping suddenly as observed here. This supports the interpretation that the observed reduction of $\rho_{edge}$ arises primarily from the activation of parasitic edge channels, while the helical channels remain mostly unaffected. More broadly, this observation indicates that, beyond suppressing bulk conduction in InAs/GaInSb based systems, dual gating and electric-field control effectively modulate parasitic edge contributions, thus addressing the second key obstacle in realizing clean topological transport. The same qualitative electric-field dependence is reproduced in a second macroscopic device (see Fig. S7 of the Supplemental Material [49]), demonstrating the reproducibility of the observed edge-transport behavior.

To evidence this, we investigate a microscopic Hall bar device under dual-gating conditions, as shown in Figure 4. An exemplary scanning electron microscopy (SEM) image of the device is provided in the inset. The length between the inner and outer contacts is 3 μm and 3.5 μm. Therefore, the edge lengths are below or close to the phase coherence length [21,59]. By sweeping $V_{TG}$ through the topological gap and tuning $V_{BG}$ as a parameter, we monitored the maximum resistance $R_{max}$ for the local configuration ($R_{03,12}$) and nonlocal configuration ($R_{01,32}$), normalized to their respective quantized value $R_q = h/2e^2$ and $h/6e^2$. The resistance in the gap remains quantized to the expected value for a large range of $V_{BG}$ = -10 V to +7 V. This quantized plateau in both configurations confirms that transport is governed by topologically protected helical edge channels, with negligible parasitic bulk or edge contributions. Importantly, the stability of the quantized resistance value over a broad range of electric fields demonstrates the intrinsic robustness of the helical edge channels against electric field perturbations, suggesting that this behavior may be a general feature of quantum spin Hall systems. Above $V_{BG} \approx +7$ V, $R_{max}$ begins to decrease, signaling the onset of additional parasitic contributions, consistent with the behavior observed in the macroscopic device in Figure 3. Notably, the ratio $R_{01,32}/R_{03,12}$ remains close to 1/3 even beyond this threshold, whereas an increasing bulk contribution would strongly suppress the nonlocal resistance relative to the local resistance. This provides further evidence that the additional conduction originates from a parasitic edge contribution rather than from the bulk. A second microscopic device, measured in the local configuration and shown in Fig. S8 of the Supplemental Material [49], exhibits the same qualitative evolution with back-gate voltage. Together with the two macroscopic devices, all four devices presented here show a consistent response to dual-gate tuning. In microscopic devices, the persistence of the quantized resistance over a broad electric-field range directly reflects the resilience of the helical edge channels. The slight shift in the onset voltage between all devices is attributed to differences in dynamic capacitances, as discussed in Fig. 2. Since the bulk remains insulating throughout the whole $V_{BG}$-range (see Fig. 3(b)), the additional conduction is attributed to the population of n-type trivial edge states at more positive $V_{BG}$. These findings conclusively show that robust and dominant helical edge transport can be realized in strained TQWs by carefully tuning the external electric field using a dual-gate configuration, which effectively suppresses both parasitic bulk and edge contributions.

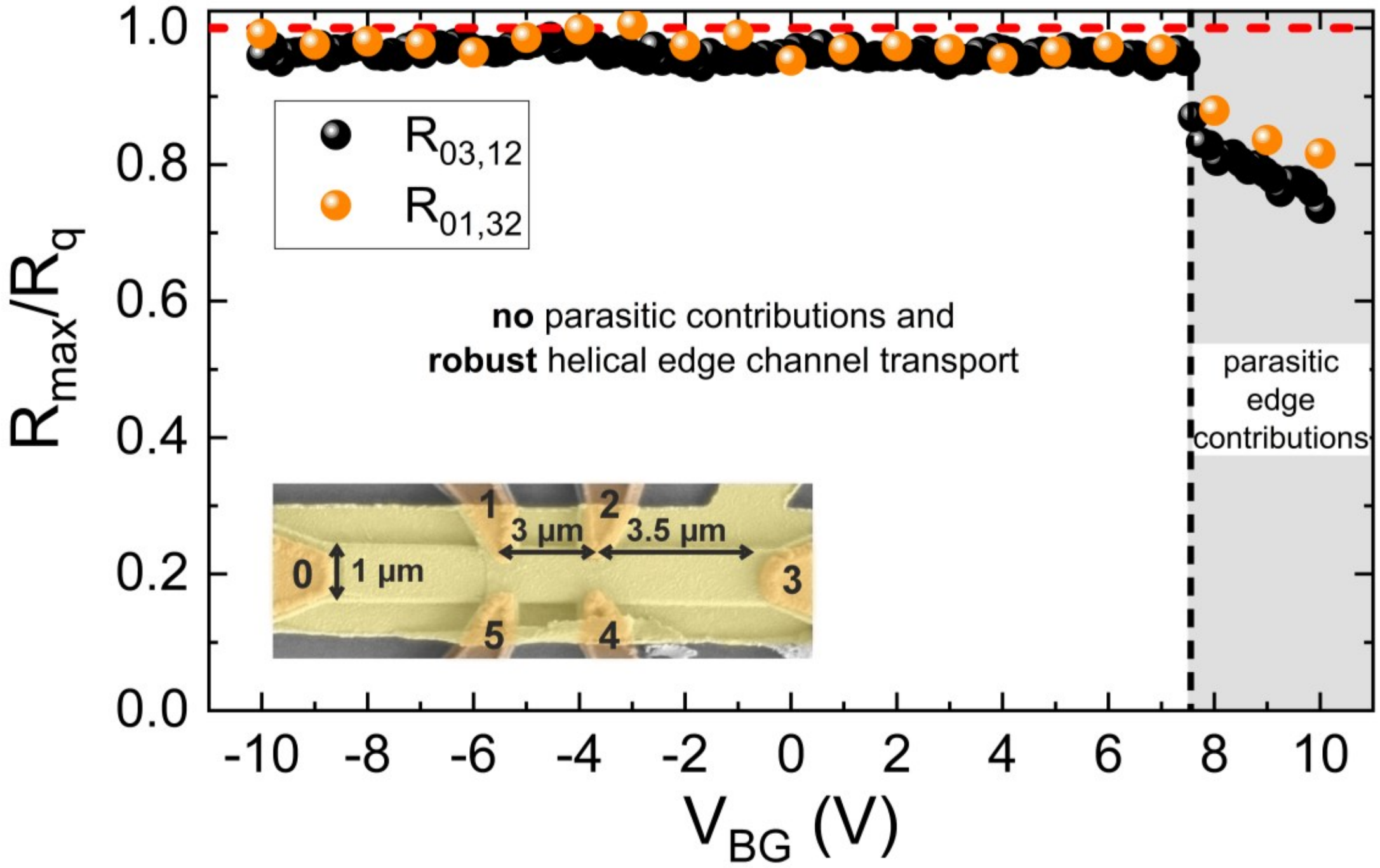

**Fig. 4.** $R_{03,12}$ (local configuration, black) and $R_{01,32}$ (nonlocal configuration, orange) normalized to their respective quantized values $h/2e^2$ and $h/6e^2$, respectively, as a function of the back-gate voltage. For $V_{BG}$ > +7 V, $R_{max}$ starts to gradually decrease in both configurations, indicating the onset of a parasitic edge contribution.

## IV. Conclusion

In conclusion, we have realized dual-gated InAs/GaInSb/InAs trilayer quantum wells grown on AlSb quasi-substrates, providing enhanced topological energy gaps and improved electrostatic control. The dual-gating scheme enables electric-field tuning of the band structure and topological gap, as evidenced by transport activation analysis and the evolution of band-structure features. Crucially, independent control of top and back gates allows simultaneous optimization of bulk insulation and edge transport, suppressing parasitic contributions and establishing dominant helical edge channel transport over a wide electric-field range. The observation of robust and quantized resistances in microscopic devices across a broad voltage space highlights the resilience of the helical edge states against electric-field perturbations. Further optimization of this architecture should enable systematic and reproducible access to the quantum spin Hall regime, while minimizing residual parasitic channels to enhance device stability and extend robust operation to higher temperatures. While the suppression of parasitic contributions relies on device- and material-specific electrostatic optimization, the observed robustness of helical edge channels against electric-field perturbations points to a more general property of quantum spin Hall systems. This intrinsic resilience is therefore expected to extend beyond the specific InAs/GaInSb platform and may be particularly attractive for device concepts based on electrostatically controlled topological transport. In contrast, the implementation of electrostatic bulk-edge optimization and the suppression of parasitic channels may depend more sensitively on the details of the material system but could provide a promising route for improving transport properties in other quantum spin Hall platforms as well.

Our results establish dual-gated InAs/GaInSb/InAs trilayer quantum wells as a scalable platform for reliably realizing and engineering the quantum spin Hall effect in III–V topological systems.

**Acknowledgments**

The work was supported by the Elite Network of Bavaria within the graduate program "Topological Insulators" and by the Occitanie region through the TOP platform and the "Quantum Technologies Key Challenge" program (TARFEP project). We acknowledge financial support from the DFG within the project HO 5194/19-1 and through the Würzburg-Dresden Cluster of Excellence on Complexity and Topology in Quantum Matter – ctd.qmat (EXC 2147, project-id 390858490). We also acknowledge the French Agence Nationale pour la Recherche (Cantor project - ANR-23-CE24-0022), Equipex+ Hybat project - ANR-21-ESRE-0026), the Occitanie region for Emergence 2024 NITRO Project, and the Physics Institutes of CNRS for Emergence 2024 - Step - project.

# Supplemental Material for ‘Electrostatic Control Enables Robust Helical Edge Channel Transport in III-V Quantum Spin Hall Insulators’

Manuel Meyer[1,a)], Tobias Fähndrich[1,2], Sebastian Schmid[1], Justus Walter[1], Martin Kamp[3], Adriana Wolf[1], Sergey Krishtopenko[1,4], Guillaume Sigu[4], Jean-Baptiste Rodriguez[5], Eric Tournie[5,6], Gerald Bastard[1,7], Frederic Teppe[4], Fabian Hartmann[1], Sven Höfling[1] and Benoit Jouault[4,b)]

[1]*Julius-Maximilians-Universität Würzburg, Physikalisches Institut and Würzburg-Dresden Cluster of Excellence ctd.qmat, Lehrstuhl für Technische Physik, Am Hubland, 97074 Würzburg, Deutschland*

[2]*Department of Physics and Astronomy, University of British Columbia, Vancouver, British Columbia, V6T1Z4, Canada*

[3] *Physikalisches Institut and Röntgen Center for Complex Material Systems, 97074 Würzburg, Germany*

[4]*Laboratoire Charles Coulomb (L2C), UMR 5221 CNRS-Université de Montpellier, Montpellier, France.*

[5]*IES, Université de Montpellier, CNRS, F-34000 Montpellier, France*

[6]*Institut Universitaire de France, F-75005 Paris, France*

[7]*Physics Department, École Normale Supérieure, PSL 24 rue Lhomond, 75005 Paris, France*

a) Email: manuel.meyer@uni-wuerzburg.de
b) Email: benoit.jouault@umontpellier.fr

## Larger section of the STEM images

A high-resolution scanning transmission electron microscopy (STEM) image covering a larger section of the heterostructure between the GaSb/AlSb superlattice (SL) and the upper AlSb barrier is shown in Fig. S1. The lower insets displays the corresponding layer structure. A GaSb/AlSb superlattice is implemented between the quasi-substrate and the active layers to reduce dislocation density and smoothen the surface [50–52]. The STEM image reveals high crystalline quality, a symmetric TQW structure and minimal surface roughness. The designed layer thicknesses for the TQW were $d_{InAs}$ = 9 nm and $d_{GaInSb}$ = 3.1 nm with an indium content of 32 % in the GaInSb layer. From the STEM image, the extracted thicknesses are $d_{InAs}$ = (8.92 ± 0.30) nm and $d_{GaInSb}$ = (3.13 ± 0.31) nm, confirming the designed values. The image confirms that all interfaces within the active TQW region are sharp, with only monolayer-scale thickness variations, indicating excellent structural uniformity. The only interface showing some non-uniformity is between the AlSb quasi-substrate and the first GaSb layer in the SL. However, this irregularity does not propagate further, as subsequent interfaces within the SL are smooth, illustrating the effectiveness of the superlattice in planarizing the surface for high-quality growth of the TQW structure. The inset on the upper-right side shows the topological band structure under zero electric field, where the gap forms between the E2 and E1 subbands with $E_{gap}$ = 27 meV.

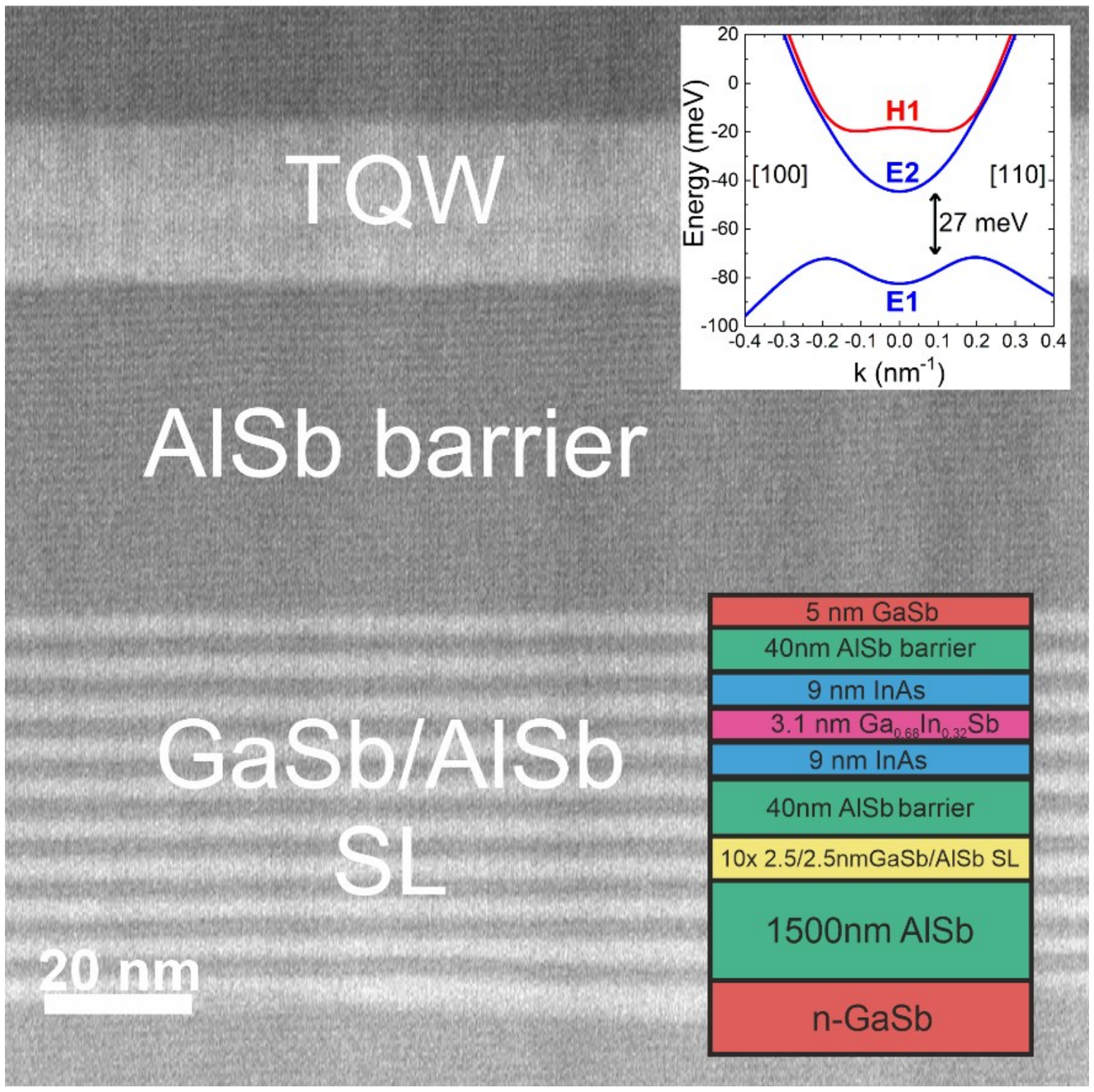

**Fig. S1.** Larger section of the STEM image from Fig. 1 from the main text. The area between the GaSb/AlSb superlattice and the upper AlSb barrier is depicted. The insets show the respective layer structures and dispersion.

## Capacitor model for TQWs

A capacitor model for a TQW as depicted in Fig. S2 similar to Ref. [13] was adapted to simulate the dynamic capacitances for different back-gate voltages.

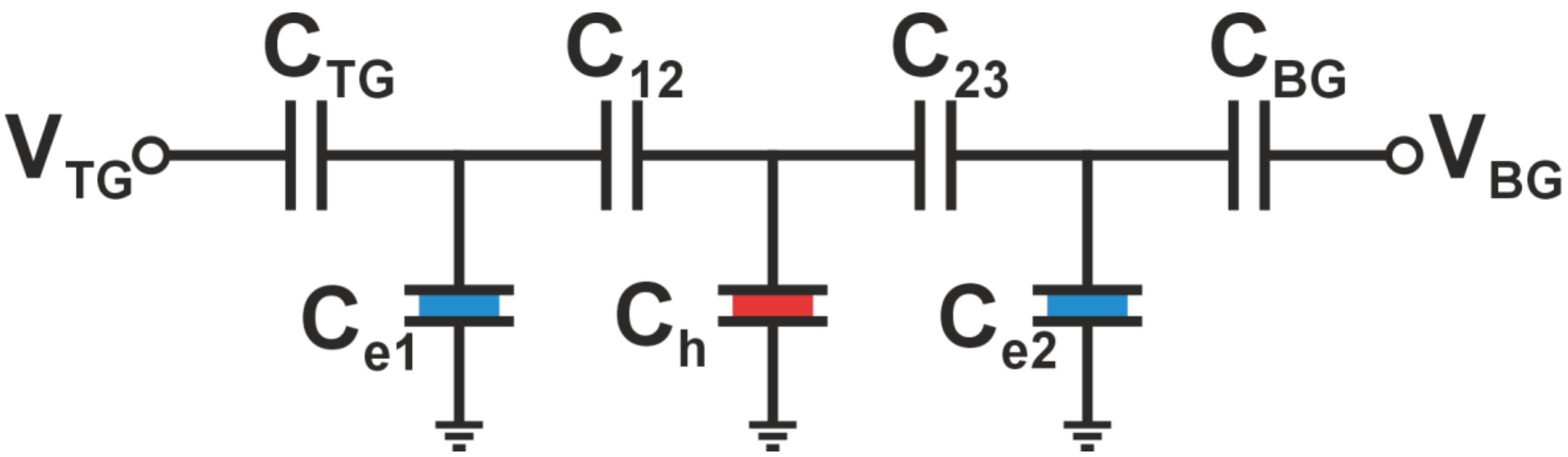

**Fig. S2.** Capacitor model for the InAs/(Ga,In)Sb/InAs trilayer quantum well device.

Here, $C_{TG}$ and $C_{BG}$ are geometric capacitances between the top-gate and back-gate and the two-dimensional electron gases (2DEGs) in the upper and lower InAs-layers, respectively. $C_{12}$ and $C_{23}$ are the geometric interlayer capacitances between both 2DEGs and the two-dimensional hole gas (2DHG) in the (Ga,In)Sb-layer. $C_{e1}$, $C_{e2}$ and $C_h$ are the quantum capacitances and are proportional to a constant 2D density of states $D = \frac{|m_{eff}|}{\pi\hbar}$, where $m_{eff}$ is the effective mass. These are $C_{e1} = C_{e2} = e^2 D_e$ and $C_h = e^2 D_h$. Thus, these can either display a finite value or zero, if the corresponding subband is filled or depleted. The geometric capacitances have been estimated with $\varepsilon_{InAs} = 15.5$, $\varepsilon_{GaInSb} = 16.4$ (with 32% In), $\varepsilon_{AlSb} = 10.9$, $\varepsilon_{SiN} = 7.0$ and $\varepsilon_{SiO2} = 3.9$ [13]. As the top gate is separated by an insulating superlattice consisting of $SiO_2$/SiN (approx. 100 nm), the upper barrier and cap layer, a constant geometric capacitance of $C_{TG} \approx 48$ nF/cm$^2$ is determined. In contrast, $C_{BG}$ is only separated by a semiconductor layer (AlSb) and the exact position at what this layer becomes insulating can vary depending on the applied electric field. To extract $C_{BG}$ as it was done in Fig. 2(a) from the main text, the slope of the top-gate position as a function of back-gate position is fitted with

$$V_{TG} = -\frac{C^2 C_2 V_{BG}}{C_{TG}(C^2 + 2CC_{BG})}, \quad (1)$$

where $C = C_{12} = C_{23} \approx 2.28\ \mu F/cm^2$. Please note that even if charges would be considered in the 2DEGs or 2DHG, the slope would not be affected, and we therefore decided not to include them for simplicity ($Ce_1 = Ce_2 = C_h = 0$). We separated three regions with three different slopes and extracted $C_{BG}$ to: $C_{1,BG} \approx$ 15 nF/cm$^2$, $C_{2,BG} \approx 6$ nF/cm$^2$, and $C_{3,BG} \approx 22$ nF/cm$^2$. From these, the effective distance between the back gate and the TQW can be estimated to: $d_{1,BG} \approx 480$ nm, $d_{2,BG} \approx 1200$ nm and $d_{3,BG} \approx 330$ nm. This indicates that the distance and therefore capacitance between the back gate and the TQW is dynamic and dependent on the applied electric field.

## Electric field tuning of the topological band structure

Despite the large separation between the back gate and the TQW, the extracted back-gate capacitance $C_{BG}$ confirms that the back gate remains effective for tuning the band structure. Fig. S3(a) shows the Hall traces $R_{03,15}$ for $V_{BG}$ = -10 V on both the conduction ($V_{TG}$-$V_{CNP}$ = +1 V and +2 V) and valence ($V_{TG}$-$V_{CNP}$ = -1 V and -2 V) band sides. While the conduction-band traces remain linear, the valence-band traces clearly exhibit two-carrier behavior. Corresponding charge carrier densities extracted using a two-carrier model [13,20] are shown in Fig. S3(b), where only electrons can be resolved in the conduction band, while both electrons and holes are extracted in the valence band. This is further supported by the presence of a van-Hove singularity in the valence band only.

In contrast, Hall traces in Fig. S3(c) for $V_{BG}$ = +6 V display two-carrier behavior in both the conduction and valence bands. The extracted carrier densities in Fig. S3(d) confirm the presence of electrons and holes on both sides of the gap, and van-Hove singularities are visible in both bands. These results clearly demonstrate the tunability of the topological band structure with the applied electric field, consistent with the band structure schematics shown in the bottom-right insets of Fig. S3(b) and (d). Note that in the insets of the resistance trace the up-sweep is depicted, as the visibility of both VHS is better due to a lower quantum scattering time [56].

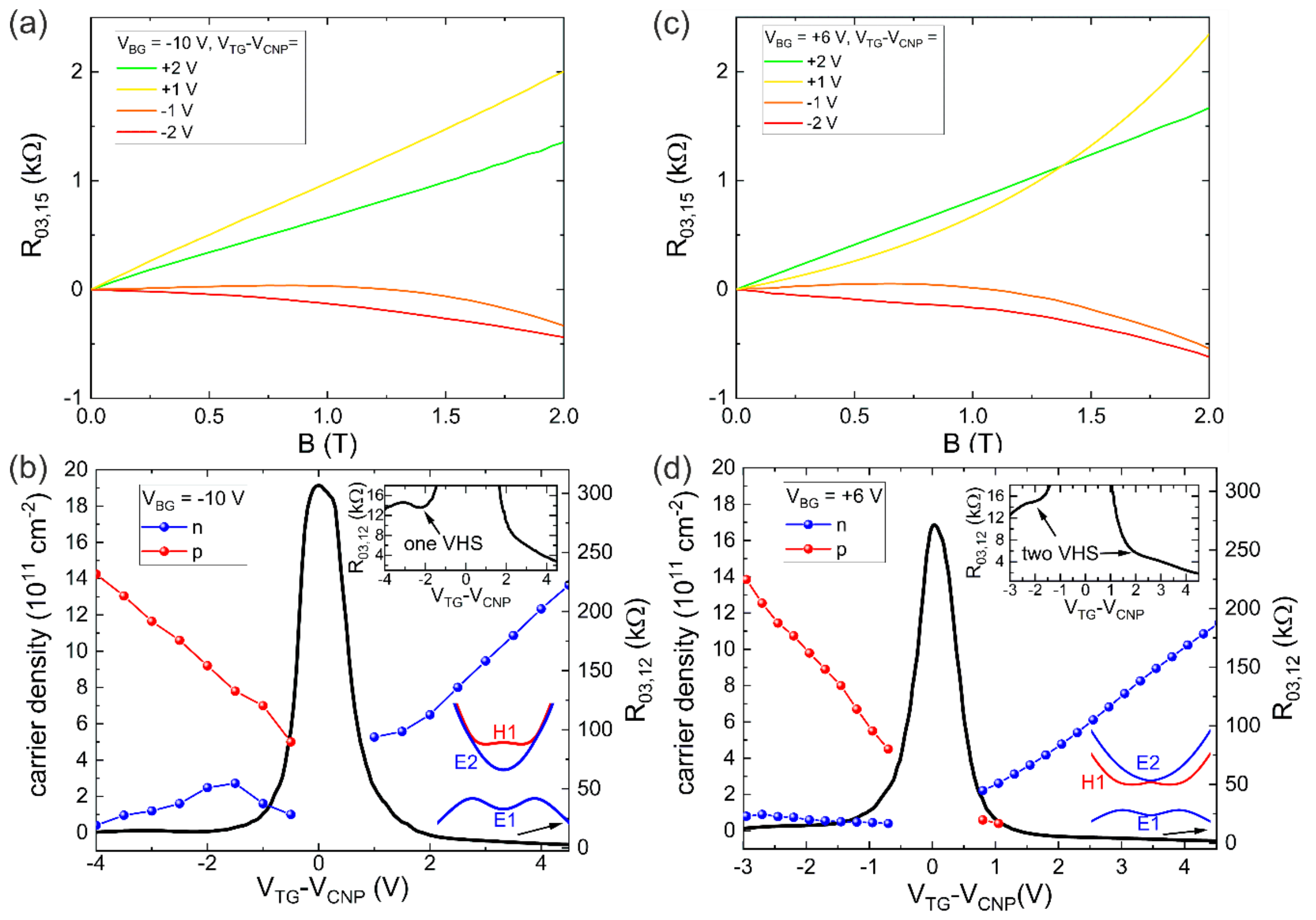

**Fig. S3.** (a) Hall traces $R_{03,15}$ for $V_{BG}$ = -10 V and (b) extracted charge carrier densities for electrons (blue) and holes (red) compared to the longitudinal resistance $R_{03,12}$ and the charge neutrality point. (c) Hall traces $R_{03,15}$ for $V_{BG}$ = +6 V and (d) extracted charge carrier densities for electrons (blue) and holes (red) compared to the longitudinal resistance $R_{03,12}$ and the charge neutrality point.

## Back-gate controlled Shubnikov–de Haas oscillations and Hall response

As for the SdH oscillations and Hall plateaus in general, exemplary magnetotransport measurements at $V_{TG}$ = +10 V are shown in Fig. S4.

In Fig. S4(a), $R_{03,12}$ and $R_{03,15}$ at $V_{BG}$ = -10 V are presented. Pronounced SdH oscillations and well-developed Hall plateaus are observed, yielding a maximum mobility of $\mu = 7.8 \times 10^4$ cm$^2$/Vs at a carrier density of $n = 2.33 \times 10^{12}$ cm$^{-2}$. The longitudinal resistance exhibits a dominant high-frequency component and a much weaker low-frequency contribution. These two frequencies are attributed to the two InAs quantum wells [20,12]. Due to the positive top-gate voltage and negative back-gate voltage, transport is dominated by the upper InAs well, while the lower well is only weakly populated.

In contrast, Fig. S4(b) shows the data for $V_{BG}$ = +6 V. Here, two distinct SdH frequencies are clearly resolved in the longitudinal resistance, indicating significant occupation of both InAs wells. The increased back-gate voltage enhances the carrier density in the lower well. A slight overall increase in $R_{03,12}$ is observed. The observed positive magnetoresistance can be consistent with the presence of two conducting channels, as evidenced by the two SdH frequencies. Nevertheless, a high maximum mobility of extracted $\mu = 6.9 \times 10^4$ cm$^2$/Vs at a carrier density of $n = 2.79 \times 10^{12}$ cm$^{-2}$. We note that the SdH-oscillations do not reach exactly zero with a small offset being present. Such small offsets in the longitudinal resistance of the SdH oscillations can be attributed to experimental uncertainties or a slight admixture of Hall voltage due to geometric asymmetries. In addition, the simultaneous occupation of both InAs quantum wells in the trilayer structure can give rise to a small residual background in the longitudinal resistance, as parallel transport contributions from two channels prevent the SdH minima from vanishing completely despite well-developed Landau quantization.

These results demonstrate that the trilayer heterostructure maintains high material quality and low disorder even when both quantum wells are substantially populated, as evidenced by the well-defined SdH oscillations and quantized Hall plateaus.

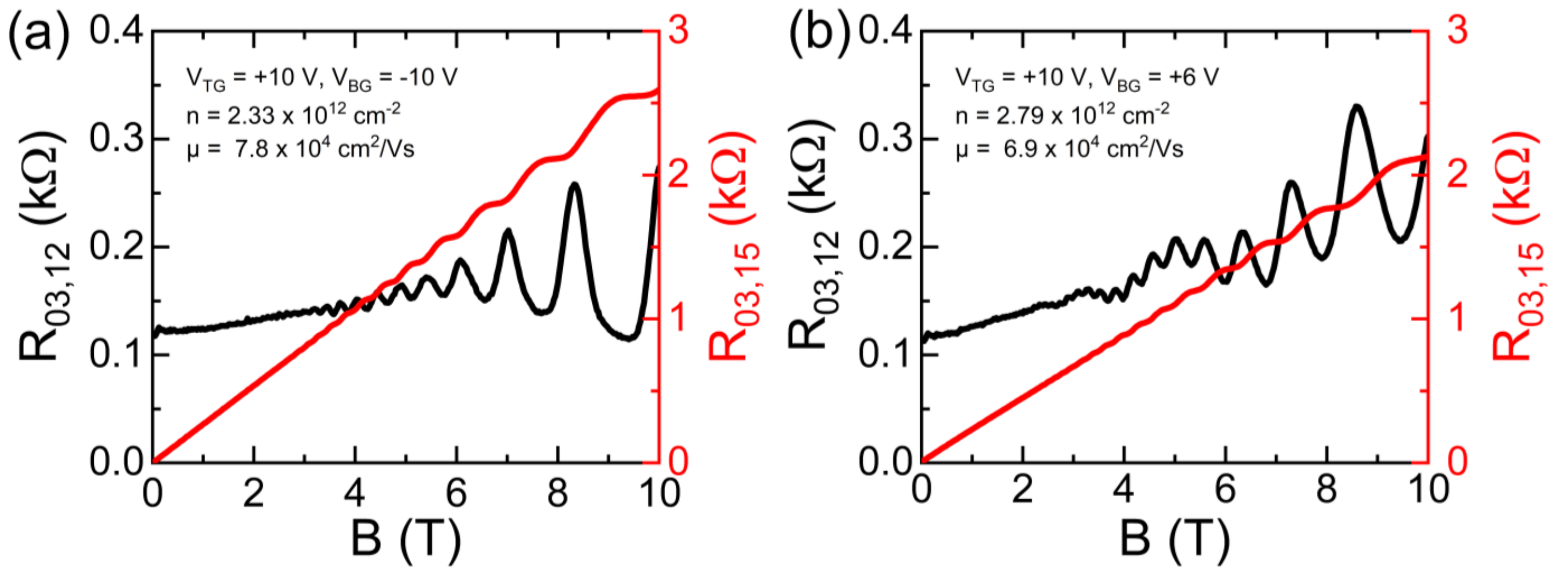

**Fig. S4.** Longitudinal ($R_{03,12}$, black) and Hall ($R_{03,15}$, red) resistance for (a) $V_{BG}$ = -10 V and (b) $V_{BG}$ = +6 V.

## Electric-field-dependent band structure calculations

Band structure calculations were performed using the eight-band k·p Hamiltonian [11], which explicitly accounts for the interactions between the Γ6, Γ8, and Γ7 bands in the bulk materials. Strain effects arising from lattice mismatch between the buffer, quantum well layers, and AlSb barriers are included in the Hamiltonian. The eight-component envelope wave functions were expanded in a plane-wave basis, and the resulting eigenvalue problem was solved numerically.

To model the effect of external electric fields and gate voltages, we self-consistently solved the Poisson equation across the heterostructure, including the dual-gate geometry. The resulting potential profile was then incorporated into the k·p Hamiltonian, allowing us to capture the electric-field-induced band shifts and modifications of band inversion. More details of the Hamiltonian can be found in Ref. [11]. All bulk material parameters and valence band offsets are taken from Ref. [11,53]. As for a qualitative comparison of the electric-field-dependent calculations with the experiments, we estimated the electric field caused by both gates at the TQW with:

$$E_j = \frac{-\Delta\varphi}{\varepsilon_j\left[\frac{L_1}{\varepsilon_1}+\frac{L_2}{\varepsilon_2}+\cdots\ldots+\frac{L_N}{\varepsilon_N}\right]}. \tag{1}$$

$E_j$ is the electric field in the jth layer (here the TQW) with dielectric constant $\varepsilon_j$, $\Delta\varphi$ is the voltage drop across the entire heterostructure from the top and back gate, and $L_N$ and $\varepsilon_N$ are the thicknesses and dielectric constants of the other layers in the stack. Using this approach, the maximum electric field in the gap at the TQW was estimated to be 200 kV/cm (for positive back gates and negative top gates), providing a benchmark for comparison with calculated band shifts.

## Additional macroscopic Hall bar

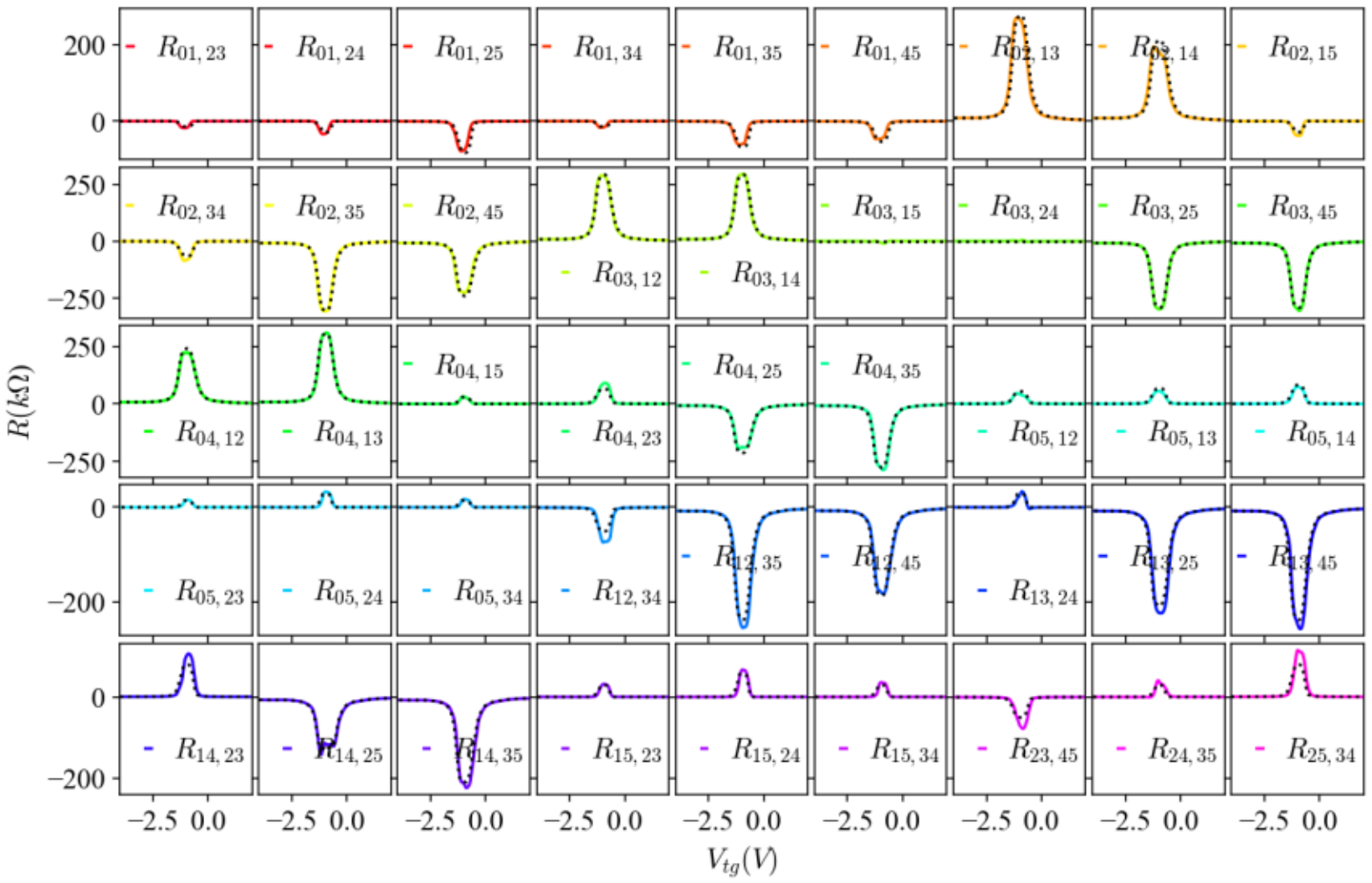

**Fig. S5.** Forty-five four-probe resistances measured on an additional macroscopic device at a bottom gate voltage $\boldsymbol{V_{BG}} = \mathbf{12\ V}$. The solid-colored lines represent the experimental data, while the dotted black lines correspond to simultaneous fits of all curves. The resistances are computed numerically using a finite-difference model with two fitting parameters: the edge and bulk conductivities. These conductivities are used to fit all 45 resistance traces simultaneously for given bottom and top gate voltages.

We present in this section results obtained on an additional macroscopic Hall bar (total length: 130 µm, width: 20 µm, distance between lateral probes: 70 µm). Figure S5 shows the 45 four-probe resistances as a function of the top gate voltage, measured at a fixed bottom gate voltage $V_{\mathrm{BG}} = 12$ V and a temperature of $T = 1.8$ K. The resistance peak associated with the inverted gap is clearly visible. The experimental data are shown as colored curves, while the corresponding multiprobe fit is indicated by a dotted line. This fit is based on a finite-difference model incorporating both edge and bulk conductivities of the Hall bar [59]. For given values of these two parameters, the four-probe resistances are computed numerically. The edge and bulk conductivities are then estimated by the best fit of the experimental resistances for each gate voltage.

The extracted edge and bulk conductivities are plotted as a function of the top gate voltage in Fig. S6a. The topological gap is evidenced by a strong suppression of the bulk conductivity (blue curve). In this regime, the bulk resistivity remains very large, corresponding to resistances up to 5 MΩ. By contrast, the residual edge conductivity in the gap is about 300 μS·μm.

The same fitting procedure can be applied at different bottom gate voltages, as shown in Fig. S6b–d for $V_{\mathrm{BG}} = 6$ V, 0 V, and −6 V, respectively. Two main conclusions can be drawn: (i) the bulk conductivity remains negligible over the entire range of bottom gate voltages; (ii) the edge conductivity is nearly constant from −6 V to 6 V, around 200 μS·μm and increases significantly at 12 V.

These results are reproducible for both sweep directions of the gate voltage. Since bulk transport is negligible, the observed increase at $V_{\mathrm{BG}} = 12$ V can be interpreted either as a sudden increase of the backscattering length from about 6 μm to 8 μm (using $\lambda = G_{\mathrm{edge}} R_K$, with $R_K = h/e^2$), hardly explainable, or, more simply, as the onset of an additional parasitic edge conduction channel.

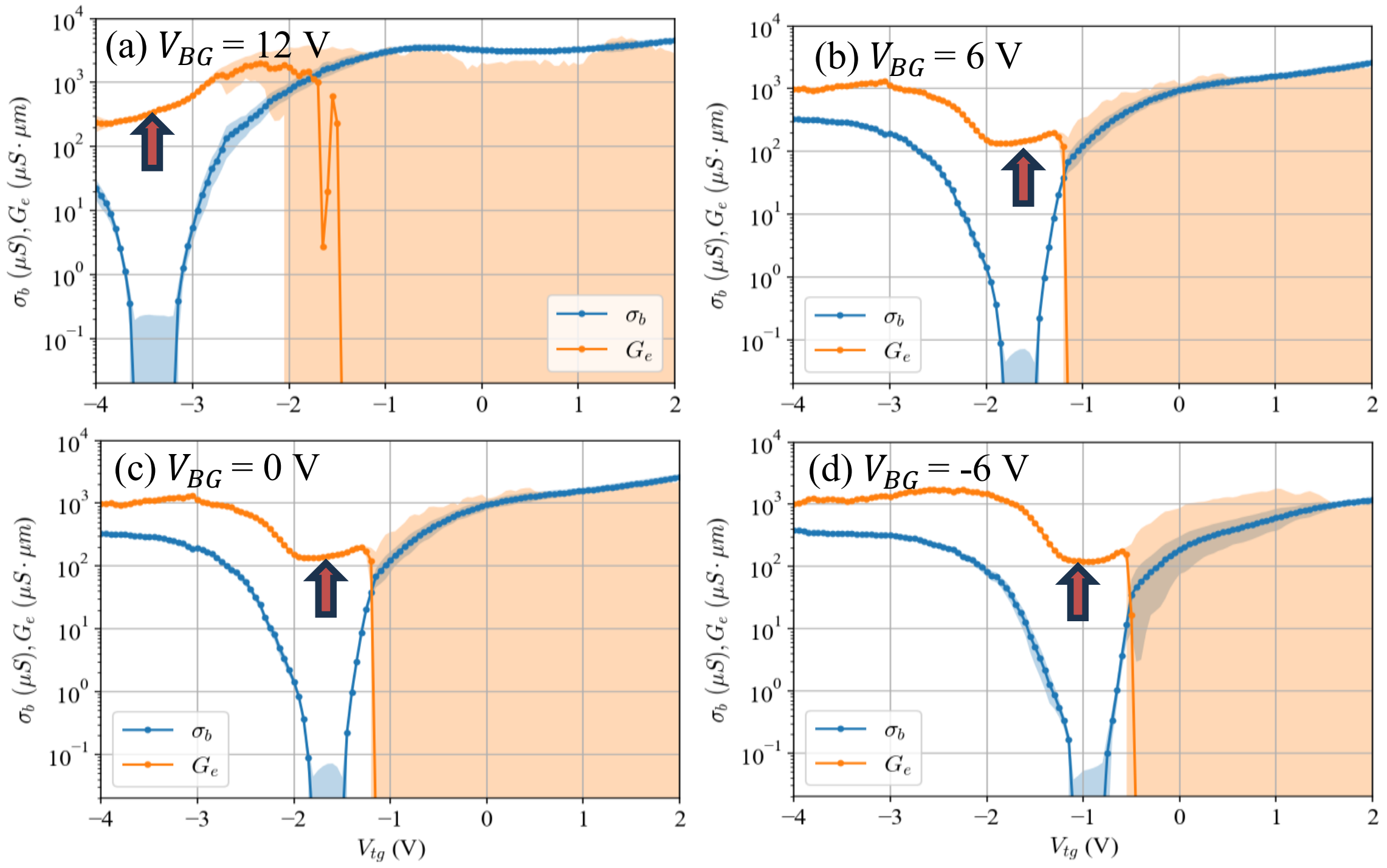

**Fig. S6.** Evolution of the fitting parameters of the four-probe resistances, as a function of the top gate voltage, for four different bottom gate voltages: (a) 12 V, (b) 6 V, (c) 0 V and (d) -6 V, for the additional macroscopic Hall bar. The bulk conductivity in the gap is below 0.1 μS, whatever the bottom gate voltage.

The edge conductivity is noticeably higher at $V_{BG}$ = 12 V. The evolution of the edge conductivity in the topological gap is indicated by large, orange arrows.

We note that a residual nonlocal response persists on the valence-band side of the resistance peak. As indicated by Fig. S3, the accessible gate range does not reach a well-defined single-carrier hole regime but remains in a regime with coexisting electron and hole contributions. Nonlocal transport persisting towards the valence-band side of the charge-neutrality region has also been reported in InAs/GaSb quantum wells [48,30,34]. The microscopic origin of the residual nonlocal signal observed here cannot be conclusively determined from the present measurements, and we therefore refrain from assigning it to a specific mechanism.

Finally, to demonstrate the reproducibility of the results presented in Fig. 3b, main text, Fig. S7 shows the edge and bulk resistivities in the gap as a function of the bottom gate voltage. As in Fig. 3b, the sudden drop in the edge resistivity at positive $V_{BG}$ is evidenced. By contrast, there is no obvious drastic change in the bulk resistivity as it remains very large in the whole $V_{BG}$ range.

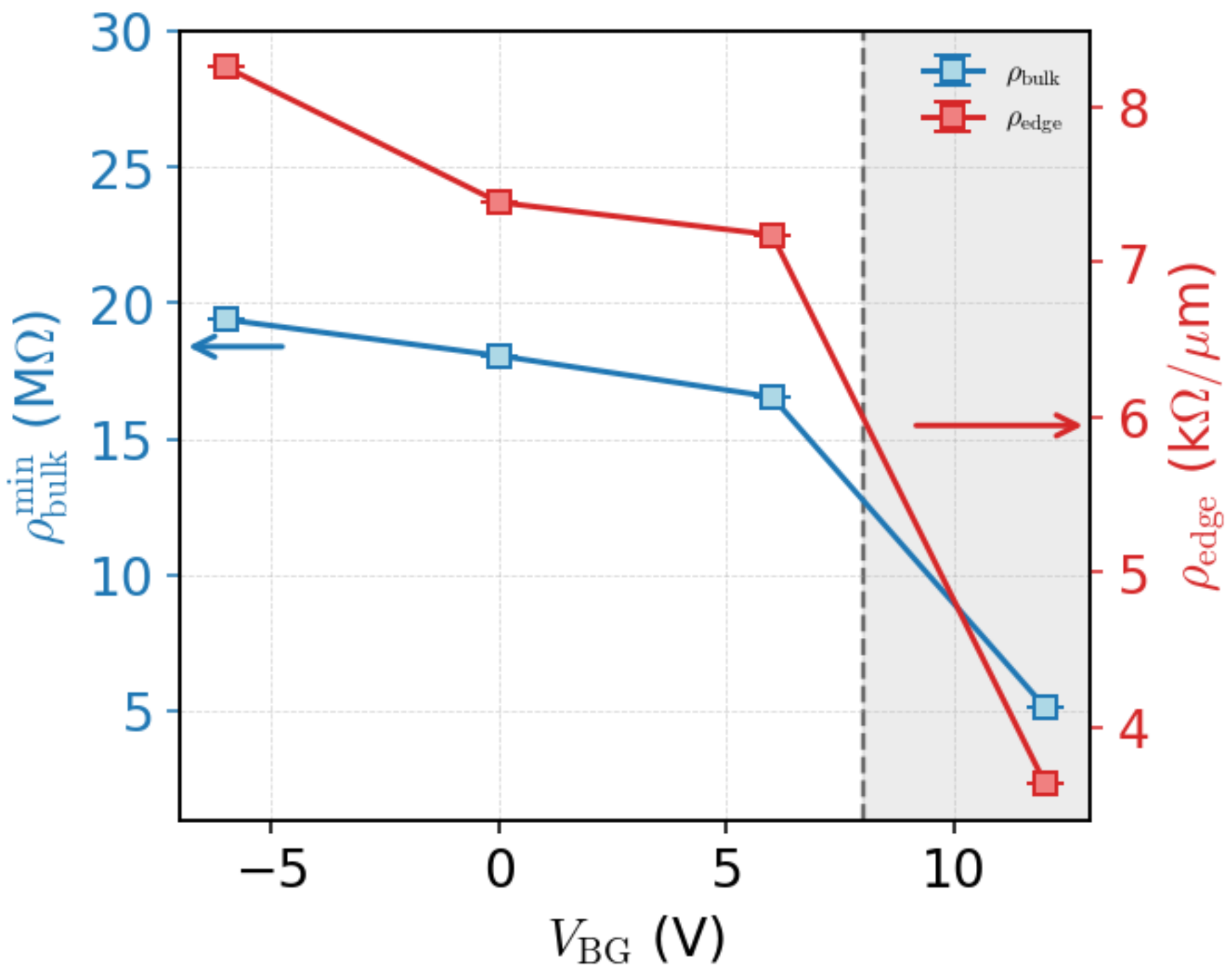

**Fig. S7.** Extracted minimal bulk resistivity $\boldsymbol{\rho_{bulk}^{min}}$ (taking into account the error bar of the fit in Fig. S6) and edge resistivity $\boldsymbol{\rho_{edge}}$ of the additional macroscopic Hall bar, based on the multi-probe model. The contribution of $\rho_{bulk}$ is very low and could not be determined precisely by the model. Only the minimal possible value $\boldsymbol{\rho_{bulk}^{min}}$ is reported. The bulk resistivity is always larger than 5 MΩ per sq, evidencing an insulating bulk. Between $\boldsymbol{V_{BG}}$ = -10 V to +6 V, ρedge is roughly constant at 7 kΩ/μm. For $\boldsymbol{V_{BG}}$ > +6 V (grey region), $\boldsymbol{\rho_{edge}}$ decreases sharply, signaling the onset of parasitic edge contributions.

## Additional microscopic Hall bar

We further investigated a second microscopic Hall bar with a length of 3 μm over a back-gate voltage range from $V_{BG}$ +10 V to -10 V in the local geometry $R_{03,12}$ in Fig. S8. It shows the same overall trend as the device from the main text, demonstrating reproducibility across devices, albeit with fewer back-gate points. The only difference is a local maximum above the quantized resistance around $V_{BG}$ = +7 V, possibly due to a decrease of the phase coherence length, as observed in previous measurements [60,21]. However, regardless of the coherence length, $R_{03,12}$ should not drop below the expected quantized value $h/2e^2$, as observed above $V_{BG}$ = +7 V, unless additional edge and/or bulk conduction is present.

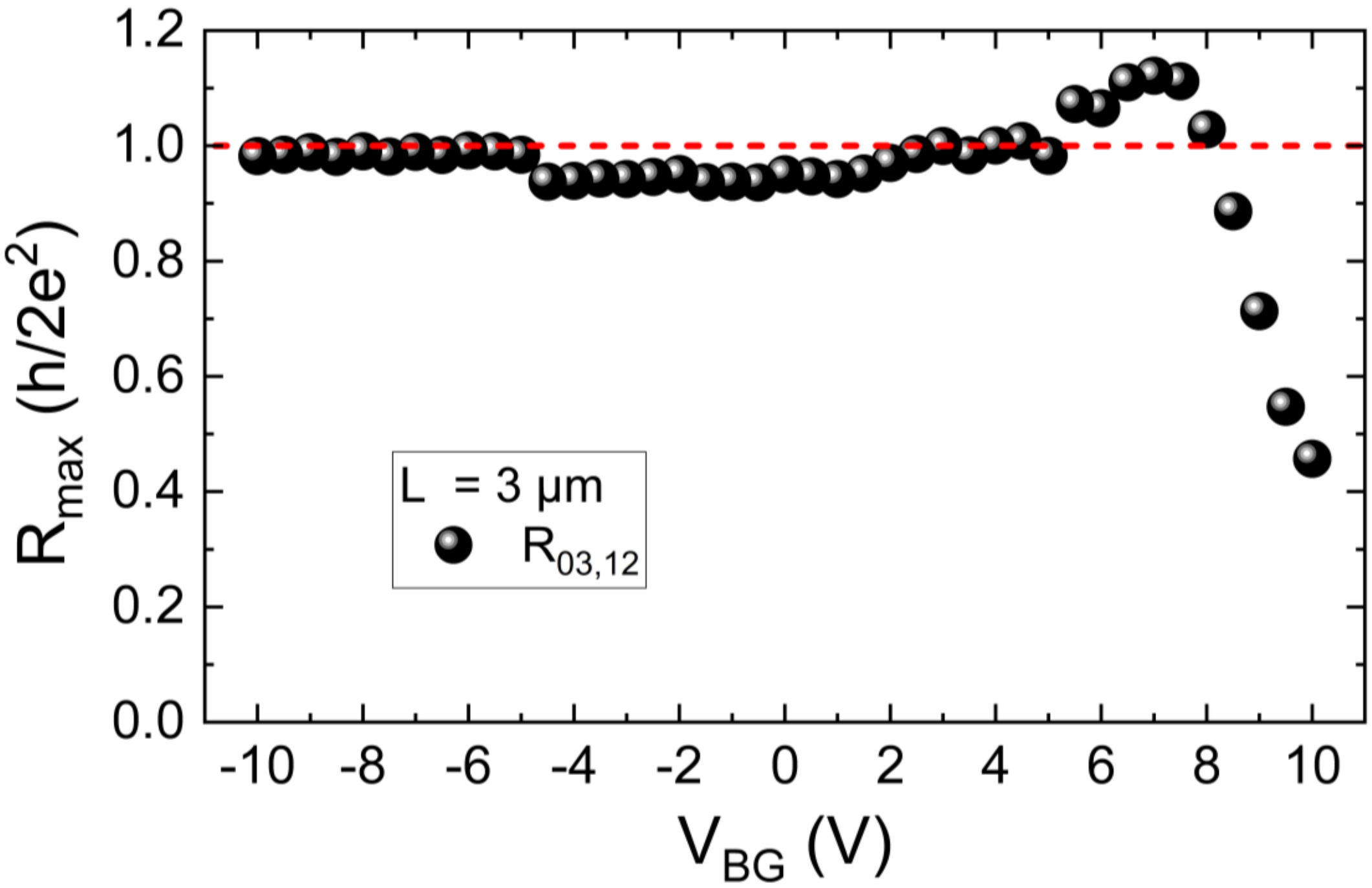

**Fig. S8.** $R_{03,12}$ (black) for another 3 μm device as a function of $V_{BG}$. The trend is qualitatively similar to the device from the main text.

## AFM scans of sample surface

To further substantiate the structural quality and uniformity of the heterostructure, we have performed additional AFM measurements on two distinct positions on the sample, each imaged at two resolutions. In Fig. S9(a), the first area is shown with a large scan (2 μm × 2 μm) and a higher-resolution inset (500 nm × 500 nm), yielding average surface roughness values of $R_a$ = 0.32 nm and $R_a$ = 0.26 nm, respectively. For the second area in Fig. S9(b), $R_a$ = 0.38 nm is extracted for the large scan and $R_a$ = 0.20 nm for the smaller scan. The three prominent white dots visible at the bottom of the large-area scan in Fig. S9(b) likely result from residual surface impurities and are not growth-related; they do not affect the overall smoothness of the TQW. Overall, these AFM images indicate minimal surface roughness and excellent sample uniformity, confirming high structural quality and ruling out disorder-related effects that could influence transport measurements.

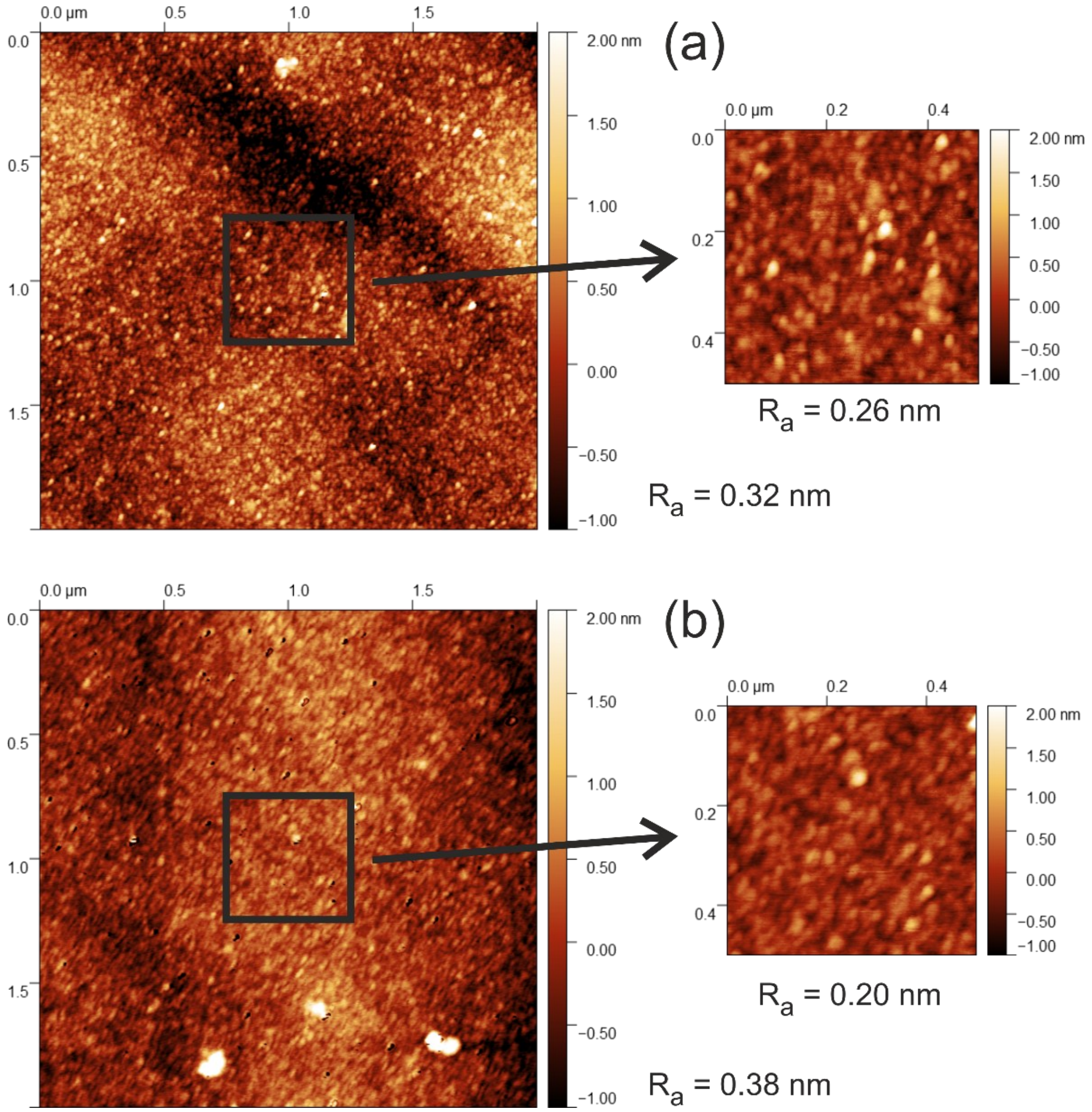

**Fig. S9.** AFM images of two different positions on the sample, shown for both a large-area scan (2 µm × 2 µm) and a smaller, higher-resolution scan (500 nm × 500 nm). The measured average surface roughness is $R_a < 0.38$ nm for all scans, demonstrating excellent structural quality and uniformity, excluding disorder-related effects.